\documentclass{article}
\usepackage{amsmath}
\usepackage{graphicx}

\input{tcilatex}

\begin{document}

\author{D. Foerster \\
CPMOH, Universit\'{e} de Bordeaux 1 \\
351, cours de la Lib\'{e}ration \\
33405 Talence Cedex, France}
\title{On an "interaction by moments" property of four center integrals.}
\maketitle

\abstract{ The four center integrals needed in the Hartree Fock approximation 
and in TDDFT linear response are known to be difficult to calculate for orbitals of the Slater type or of finite range.  
We show that the interaction of pairs of products that do not mutually intersect may be replaced 
by the interaction of their moments, of which there are O(N). Only quadruplets  of  
orbitals 'close' to one another need an explicit calculation and the total calculational 
effort therefore scales as O(N).  We provide a new and concise proof of this "interaction by moments" property. }%

\section*{Motivation}

This note is motivated by the occurrence of four center integrals in the
TDDFT linear response equation \cite{Petersilka}, \cite{Casida} 
\begin{equation}
\chi =\frac{\delta \rho (r,t)}{\delta V_{ext}(r^{\prime },t^{\prime })}=%
\frac{1}{\chi _{KS}^{-1}-\left( \frac{\delta (t-t^{\prime })}{|r-r^{\prime }|%
}+\frac{\delta V_{xc}(r,t)}{\delta \rho (r^{\prime }t^{\prime })}\right) }
\label{linear_response}
\end{equation}%
where $\chi _{KS}=\frac{\delta \rho }{\delta V_{KS}}$ and $\chi =\frac{%
\delta \rho }{\delta V_{ext}}$ are the free and interacting density
response, where $\rho $ denotes the electronic density and where $V_{ext}$
is an external potential acting on the electrons. In a basis of local
orbitals $\{f^{a}(r)\}$, the electronic density $\rho $ and the external
potential $V_{ext}$ may be expanded in terms of products of such orbitals
according to

\begin{equation}
\rho =\sum_{a,b}f^{a\ast }f^{b}\rho _{ab}\text{ , \ }V_{ext}^{ab}=\int
drV_{ext}f^{a\ast }f^{b}
\end{equation}%
Equation (\ref{linear_response}) then turns into a matrix equation for the
response $\chi _{ab,cd}=\frac{\delta \rho _{ab}}{\delta V^{cd}}$ of $\rho
_{ab}(t)$ with respect to variations of $V_{ext}^{cd}(t^{\prime })$ and this
equation then contains the Coulomb interaction between products of orbitals
or 'four center integrals' 
\begin{equation}
<12|\frac{1}{r}|34>=\int drdr^{\prime }f^{1\ast }(r)f^{2}(r)\frac{1}{%
|r-r^{\prime }|}f^{3\ast }(r^{\prime })f^{4}(r^{\prime })
\end{equation}%
A technique for calculating these quantities in terms of two center
integrals was developed in \cite{Talman84}. In an alternative "resolution of
identity" method, products of orbitals are replaced by auxiliary functions,
see \cite{ResolutionIdentity}, \cite{Gisbergen}. \ 

For orbitals of finite range, there are, for $N$ atoms, $O(N^{2})$ pairs $%
(12)$, $(34)$ of individually intersecting orbitals suggesting the need of $%
O(N^{2})$ distinct calculations. Here we show that only the subset of $<12|%
\frac{1}{r}|34>$ $\ $where a pair $(1,2)$ intersects with a pair $(3,4)$
must be calculated explicitly, while the remaining ones can be taken into
account by their multipolar interactions. Since there are only $O(N)$ such
quadruplets of orbitals and because the effort of calculating the moments
scales like $N$, the cost of calculating four center integrals then scales
as $O(N)$.

The present note arose in an ongoing effort to implement linear response for
extended molecular systems, an effort prompted by recent work \cite%
{BlaseOrdejon}. We prove and exploit an observation of Greengard on the
exact character of the "interaction by moments". This observation was
rederived previously in the literature \cite{HeadGordon} and its consequence
has recently been absorbed in a corresponding computer code \cite{Hogekamp},
but our concise and simple deduction of this important property may still be
of interest.

\section*{Reduction from four centers to two centers}

We first need a reduction of products of orbitals to a set of single center
functions. Following the discussion of \cite{Talman84} we obtain an
expansion in spherical harmonics of a translated function $f(\overrightarrow{%
r}-\overrightarrow{a})$ by using its momentum representation 
\begin{eqnarray}
\psi _{lm}(\overrightarrow{r}-\overrightarrow{a}) &=&\int \frac{d^{3}p}{%
(2\pi )^{3/2}}\psi _{lm}(\overrightarrow{p})e^{-ip(r-a)}\text{ \ }
\label{translation} \\
\psi _{lm}(\overrightarrow{p}) &=&i^{l}\widetilde{\psi }_{l}(p)Y_{lm}(%
\overrightarrow{p})\text{, }\widetilde{\psi }_{l}(p)=\sqrt{\frac{2}{\pi }}%
\int_{0}^{\infty }r^{2}\psi _{l}(r)j_{l}(pr)dr  \notag
\end{eqnarray}%
We use spherical Bessel functions $j_{l}(x)$ with $l$ integer, that are
related to conventional Bessel functions for half integers by $j_{l}(x)=%
\sqrt{\frac{\pi }{2x}}J_{l+\frac{1}{2}}(x)$. In practice, fast Hankel
transform routines \cite{FastHankel} are needed to speed up the calculation.
Expanding $e^{i\overrightarrow{p}\cdot \overrightarrow{r}}$ in spherical
waves, one finds%
\begin{equation}
\psi _{lm}(\overrightarrow{r}-\overrightarrow{a})=%
\sum_{l_{1}m_{1}}Y_{l_{1}m_{1}}(\overrightarrow{r})G_{l_{1}m_{1}}(r)
\label{translated_1}
\end{equation}%
where the coefficients $G_{l_{1}m_{1}}(r)$ that multiply the spherical
harmonics $Y_{l_{1}m_{1}}(\overrightarrow{r})$ now depend on both quantum
numbers $l_{1}$ and $m_{1}$ because spherical symmetry is lost. One has the
following expression for $G_{l_{1}m_{1}}(r)$

\begin{eqnarray}
G_{l_{1}m_{1}}(r) &=&2\sqrt{8\pi }\sum_{l_{2}}F_{l\text{ }%
l_{1}l_{2}}(r)Y_{l_{2},m_{1}-m}^{\ast }(\overrightarrow{a})(-)^{l_{1}}(-)^{%
\frac{1}{2}(l_{2}+l_{1}+l)}G_{lm,l_{1}m_{1}l_{2}m_{1}-m}\text{ \ \ } \\
F_{l\text{ }l_{1},l_{2}}(r) &=&\int_{0}^{\infty }j_{l_{1}}(pr)\widetilde{%
\psi }_{l}(p)j_{l_{2}}(pa)p^{2}dp  \notag
\end{eqnarray}%
where $G_{l_{1}m_{1},l_{2}m_{2}l_{3}m_{3}}$ are Gaunt coefficients for the
overlap of three spherical harmonics. To bring this approach into
perspective, it is interesting to consider an orbital with a cusp
singularity such as $e^{-r\text{ }}$and to study the convergence of the
translated orbital towards a translated cusp with increasing angular
momentum cutoff $\ j_{\max }$, see the figure. The figure shows that a
fairly large number of angular harmonics is required for the representation
of orbitals having such a cusp.

\begin{figure}
\begin{center}
\includegraphics[width=3.3466in]{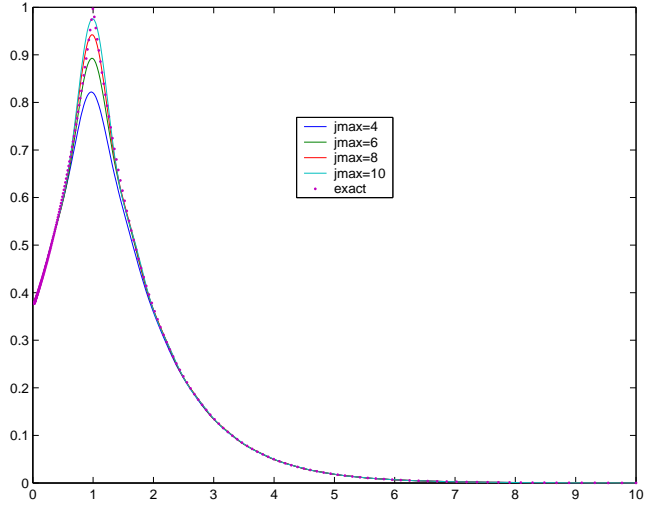}
\caption{convergence towards a cusp with $j_{\max }$}
\end{center}
\end{figure}

By applying translations to two overlapping orbitals we may obtain
an expansion of their product about a common midpoint 
\begin{equation}
\psi _{l_{1}m_{1}}(\overrightarrow{r}-\overrightarrow{a}_{1})\psi
_{l_{2}m_{2}}(\overrightarrow{r}-\overrightarrow{a}_{2})=\sum_{l=0..l_{\max
},\text{ \ \ }m=-l..l}G_{lm}(r)Y_{lm}(\overrightarrow{r})
\end{equation}%
Using such expressions we can compute the Coulomb interaction of two pairs
of mutually overlapping orbitals from the point of view of their associated
effective centers.

\section*{Exact interaction by moments of non intersecting centers}

We have seen that pairs of orbitals may be replaced by effective centers.
Now we wish to show that the interaction between effective centers may be
simplified when their spheres of support no longer intersect. We begin by
quoting formulas for the computation of two center integrals in Fourier
space, see \cite{Talman84}, \cite{siestaSoler}

\begin{equation}
<1|\frac{1}{r}|2>=\int \psi _{l_{1}m_{1}}^{\ast }(r_{1})\frac{1}{%
|r_{1}-r_{2}|}\psi _{l_{2}m_{2}}(r_{2})d^{3}r_{1}d^{3}r_{2}=4\pi \int 
\widetilde{\psi }_{l_{1}m_{1}}^{\ast }(\overrightarrow{p})\frac{1}{p^{2}}%
\widetilde{\psi }_{l_{2}m_{2}}(\overrightarrow{p})e^{i\overrightarrow{p}%
\cdot \overrightarrow{R}}d^{3}p\text{\ }  \notag
\end{equation}%
where we used that $\frac{1}{|r_{1}-r_{2}|}=-\frac{4\pi }{\bigtriangleup }$, 
$\bigtriangleup $ being the Laplace operator. Expanding $e^{i\overrightarrow{%
p}\cdot \overrightarrow{r}}$ in spherical waves one finds 
\begin{equation}
\int \psi _{l_{1}m_{1}}^{\ast }(r_{1})\frac{1}{|r_{1}-r_{2}|}\psi
_{l_{2}m_{2}}(r_{2})d^{3}r_{1}d^{3}r_{2}=\sum_{lm}Y_{lm}(\overrightarrow{R}%
)C_{lm}^{l_{1}m_{1},l_{2}m_{2}}(R)\text{\ }  \label{expansion}
\end{equation}%
with 
\begin{eqnarray}
C_{lm}^{l_{1}m_{1},l_{2}m_{2}}(R)
&=&C_{l_{1}l_{2}l}(R)G_{l_{1}m_{1},l_{2}m_{2},lm}  \label{central_formula} \\
C_{l_{1}l_{2}l}(R) &=&(4\pi )^{2}(-)^{\frac{1}{2}(l+l_{2}-l_{1})}\int_{0}^{%
\infty }\widetilde{\psi }_{l_{1}}^{\ast }(p)\widetilde{\psi }%
_{l_{2}}(p)j_{l}(pR)d  \notag
\end{eqnarray}%
where $G_{l_{1}m_{1},l_{2}m_{2},lm}$ are the previously encountered Gaunt
coefficients and $C_{l_{1}l_{2}l}(R)$ are Wigner-Eckart like couplings, see 
\cite{Talman84}, \cite{siestaSoler} for details of the derivation. We use eq(%
\ref{translation}) to rewrite the $C_{l_{1},l_{2},l}(R)$ in terms of the
original radial wave functions as follows:%
\begin{equation}
C_{l_{1},l_{2},l}=32\pi (-)^{\frac{l+l_{2}-l_{1}}{2}}\int
dr_{1}dr_{2}r_{1}^{2}\widetilde{\psi }_{l_{1}}^{\ast }(r_{1})r_{2}^{2}%
\widetilde{\psi }_{l_{2}}(r)\int_{0}^{\infty
}j_{l_{1}}(pr_{1})j_{l_{2}}(pr_{2})j_{l}(pR)dp  \label{hollow_sphere}
\end{equation}%
The "interaction by moments" property we are after is contained in the
following integral of Bessel functions 
\begin{equation}
I_{_{l_{1},l_{2},l}}(r_{1},r_{2},R)=\int_{0}^{\infty
}j_{l_{1}}(pr_{1})j_{l_{2}}(pr_{2})j_{l}(pR)dp
\end{equation}%
The Coulomb interaction coefficients $C_{l_{1},l_{2},l}$ would reduce to $%
I_{_{l_{1},l_{2},l}}(r_{1},r_{2},R)$ if the original orbitals functions were
concentrated at, respectively, radii $r_{1}$ and $r_{2}$. This integral
therefore represents the Coulomb interaction of two charged hollow shells of
radii $r_{1}$, $r_{2}$ at a distance of $R$, and with the charge densities
having the appropriate multipolar angular dependences. For $R>r_{1}+r_{2}$
where these shells no longer intersect, we expect their interaction to
simplify. Because of the Gaunt coefficients in eq(\ref{central_formula}) we
only need this interaction for even values of $l_{1}+l_{2}+l$ and where a
triangle inequality $|l_{1}-l_{2}|\leq l\leq l_{1}+l_{2}$ holds. In this
case the integrand associated with the $dp$ integration in eq(\ref%
{hollow_sphere}) is symmetric as a function of $p$ and we may therefore
extend the domain of integration to the entire $p$ axis: 
\begin{equation*}
I_{_{l_{1},l_{2},l}}(r_{1},r_{2},R)=\frac{1}{2}\int_{-\infty }^{\infty
}j_{l_{1}}(r_{1}p)j_{l_{2}}(r_{2}p)j_{l}(Rp)dp\text{ \ for }l_{1}+l_{2}+l%
\text{ even }
\end{equation*}%
It is convenient to use $j_{l}(z)=\func{Re}h_{l}(z)$ and to consider a
corresponding complex integral $I_{_{l_{1},l_{2},l}}^{c}$ with $%
I_{_{l_{1},l_{2},l}}=\func{Re}I_{_{l_{1},l_{2},l}}^{c}$ that involves the
Hankel function $h_{l}(z)$ : 
\begin{equation}
I_{_{l_{1},l_{2},l}}^{c}(r_{1},r_{2},R)=\frac{1}{2}\int_{-\infty }^{\infty
}j_{l_{1}}(r_{1}p)j_{l_{2}}(r_{2}p)h_{l}(Rp)dp\text{, \ }  \label{auxiliary}
\end{equation}%
For $R>r_{1}+r_{2}$ the contour of integration in $%
I_{_{l_{1},l_{2},l}}^{c}(r_{1},r_{2},R)$ can be be closed at infinity in
view of the relation 
\begin{equation}
h_{n}(p)=(-)^{n}p^{n}\left( \frac{d}{pdp}\right) ^{n}\frac{-ie^{ip}}{p}
\label{generic_definition}
\end{equation}%
Clearly, the exponential factor $e^{iRp}$from $j_{l}(Rp)$ dominates, for $%
R>r_{1}+r_{2}$, the factors $e^{\pm ipr_{1}}e^{\pm ipr_{2}}$ that arise in
the integrand in eq(\ref{auxiliary}) from the product $%
j_{n_{1}}(r_{1}p)j_{n_{2}}(r_{2}p)$. Since the integrand in $%
I_{_{l_{1},l_{2},l}}^{c}$ is analytic, except for possible singularities at $%
p=0$ and since the contour of integration can be closed in the upper half
plane, a non zero contribution to $I_{_{l_{1},l_{2},l}}^{c}$can only be due
to a residue at $p=0$. From eq(\ref{generic_definition}) the most singular
term in $h_{l}(Rp)$ at $p=0$ is%
\begin{equation*}
h_{l}(Rp)=-i(2l-1)!!\frac{e^{iRp}}{(Rp)^{l+1}}+O(p^{-l})
\end{equation*}%
Because of $j_{l_{1}}(r_{1}p)\sim \frac{(r_{1}p)^{l_{1}}}{(2l_{1}+1)!!}$ and
an analogous relation for $j_{l_{2}}(r_{2}p)$ a non zero residue is
impossible in eq(\ref{auxiliary}) unless $l$ attains the maximal value
permitted by the triangle inequality, $l=l_{1}+l_{2}$. When setting $%
l=l_{1}+l_{2}$ and closing the contour of integration in eq(\ref{auxiliary})
at infinity, there is a term $\sim 1/p$ that provides a non zero result by
elementary contour integration. Rewriting the result in terms of
conventional $\Gamma $ functions, one then finds, for $l+l_{1}+l_{2}$ \ even
and with $0\leq l\leq l_{1}+l_{2}$, the following simple result 
\begin{equation}
I_{_{l_{1},l_{2},l}}(r_{1},r_{2},R)=\delta _{l,l_{1}+l_{2}}\frac{\pi ^{3/2}}{%
8}\frac{r_{1}^{l_{1}}r_{2}^{l_{2}}}{R^{l+1}}\frac{\Gamma (l+1/2)}{\Gamma
(l_{1}+3/2)\Gamma (l_{2}+3/2)}  \label{contour_result}
\end{equation}%
Applied to the Coulomb interaction coefficients of eq(\ref{hollow_sphere})
one concludes 
\begin{eqnarray}
C_{l_{1},l_{2},l} &=&(-)^{\frac{l+l_{2}-l_{1}}{2}}\frac{\pi ^{1/2}}{%
2^{l_{1}+l_{2}-l+2}}\frac{\rho _{l_{1}}^{\ast }\rho _{l_{2}}}{R^{l+1}}\frac{%
\Gamma (l+1/2)}{\Gamma (l_{2}+3/2)\Gamma (l_{1}+3/2)}\text{ }\delta
_{l,l_{1}+l_{2}}\text{\ }  \label{multipoles} \\
\rho _{l_{1,2}} &=&4\pi \int_{0}^{r_{1,2}}drr^{2+l_{1,2}}\widetilde{\psi }%
_{l_{1,2}}(r)\text{ \ for }R\geq r_{1}+r_{2}  \notag
\end{eqnarray}%
This last equation shows very clearly that non overlapping orbitals interact
exactly via their moments, as shown first by Greengard \cite{Greengard}. For
another proof, see \cite{HeadGordon}.

\section*{Conclusion}

We conclude that only the subset of four center integrals $<12|\frac{1}{r}%
|34>$ of "close" pairs where $(1,2)$ intersects $(3,4)$ must be calculated
explicitly. Because there are only $O(N)$ such pairs of orbitals for a
system of $N$ atoms and because the multipoles are associated with only $%
O(N) $ products, the calculational effort scales as $O(N)$.

The conclusion that Coulomb integrals should be divided into far and near
field ones has already been incorporated in a quantum chemistry code \cite%
{Hogekamp}. But our derivation of the "interaction by moments property" of
four center integrals from a plain integral of a product of three spherical
Bessel functions is the simplest and most concise proof of this property
that is available.

\textbf{Acknowledgements}

It is a pleasure to thank James Talman from the University of Western
Ontario, Canada, for continued correspondence and for kindly providing a
computer code of his Hankel transform algorithm.

Useful comments by Xavier Blase (Lyon), Daniel Sanchez (San Sebastian) and
Andrei Postnikov (Metz) and discussions with the quantum chemistry group of
Ross Brown(Pau) and Isabelle Baraille (Pau) are gratefully acknowledged.

\textbf{Figure caption:} Translation away from the origin of $\ e^{-r}$ by
one unit, with $j\leq j_{\max }=4,6,8,10$ and the convergence of the result
towards a cusp with increasing angular momentum cutoff $\ j_{\max }$.

\end{document}